\documentclass{aa}

\usepackage[varg]{txfonts}

\usepackage{amsmath}
\usepackage{amssymb}
\usepackage{graphicx}
\usepackage{xcolor}
\usepackage{subfigure}
\bibpunct{(}{)}{;}{a}{}{,}
\date{}

\newcommand{\PD}[2]{\partial_{#2} #1}

\newcommand{\vet}[1]{{\mbox{\boldmath${\mathrm{#1}}$} }}  
\newcommand{\Div}{\nabla\cdot}
\newcommand{\DIV}[1]{\ensuremath{\nabla\cdot {#1}}  }
\newcommand{\GRAD}[1]{\ensuremath{\nabla {#1}}  }

\newcommand{\br}{\vet r}

\newcommand{\bv}{\vet v}

\newcommand{\rmR}{{{\mathrm{R}}}}

\newcommand{\mH}{{\mathcal{H}}}

\newcommand{\eg}{{{e.g.}}}
\newcommand{\ie}{{{i.e.}}}
\newcommand{\be}{{\mbox{\boldmath${e}$} }}
\newcommand{\bxi}{{\mbox{\boldmath${\xi}$} }}

\newcommand{\Ylm}{{\mbox{$Y_\ell^m$}}}

\newcommand{\mHz}{{\mbox{$\mathrm{mHz}$}}}
\newcommand{\days}{{\mbox{$\mathrm{days}$}}}
\newcommand{\Hz}{{\mbox{$\mathrm{Hz}$}}}
\newcommand{\df}{\mathrm{d}}

\begin{document}

\title{Simulating acoustic waves in spotted stars}
\author{Emanuele Papini\inst{\ref{inst1}}
  \and Aaron C. Birch\inst{\ref{inst1}}
  \and Laurent Gizon\inst{\ref{inst1},\ref{inst2}} 
  \and Shravan M. Hanasoge\inst{\ref{inst1},\ref{inst3}}}

\institute{Max-Planck-Institut f\"{u}r Sonnensystemforschung.
Justus-von-Liebig-Weg 3, 37077 G\"{o}ttingen, Germany\label{inst1}
\and Instit\"{u}t f\"{u}r Astrophysik, Georg-August-Universit\"{a}t G\"{o}ttingen, 37077 G\"{o}ttingen, Germany\label{inst2}
\and Tata Institute of Fundamental Research, Mumbai 400005, India\label{inst3}}

\date{Received 9 February 2015 / Accepted 19 March 2015}

\abstract
{Acoustic modes of oscillation are affected by stellar activity, however it
is unclear how starspots contribute to these changes. Here we investigate the nonmagnetic effects of starspots on global modes with angular degree $\ell \leq 2$ in highly active stars, and characterize the spot seismic signature on synthetic light curves. We perform 3D time-domain simulations of linear acoustic waves to study their interaction with a model starspot. We model the spot as a 3D change in the sound speed stratification with respect to a convectively stable stellar background, built from solar Model S. We perform a parametric study by considering different depths and perturbation amplitudes. Exact numerical simulations allow the investigation of the wavefield-spot interaction beyond first order perturbation theory. The interaction of the axisymmetric modes with the starspot is strongly nonlinear. As mode frequency increases, the frequency shifts for radial modes exceed the value predicted by linear theory, while the shifts for the $\ell=2, m=0$ modes are smaller than predicted by linear theory, with avoided-crossing-like patterns forming between the $m=0$ and $m=1$ mode frequencies. The nonlinear behavior increases with increasing spot amplitude and/or decreasing depth. Linear theory still reproduces the correct shifts for nonaxisymmetric modes. In the nonlinear regime the mode eigenfunctions are not pure spherical harmonics, but rather a mixture of different spherical harmonics. This mode mixing, together with the frequency changes, may lead to misidentification of the modes in the observed acoustic power spectra.}


\maketitle

\section{Introduction}

\subsection{Motivation: Low degree stellar oscillations}

Starspots are the main observed features of magnetic activity in stars and play a fundamental role in understanding stellar dynamos. 
They have been detected in many stars through the modulation of light curves \citep{Mosser2009} and Doppler imaging \citep{Strassmeier2009}, the latter suggesting polar and high- to mid-latitude concentrations of magnetic fields.
\citet{Garcia2010} unambiguously detected an activity cycle in a solar-like star with asteroseismology for the first time, using data collected by the \textit{Convection, Rotation \& planetary Transits} (CoRoT) mission \citep{Corot2009}. Other excellent data are available from the \textit{Kepler} \citep{borucki2010} satellite and even more will come from the upcoming missions \textit{Transiting Exoplanet Survey Satellite} (TESS) to be launched in 2017 \citep{tess2014}, and \textit{PLAnetary Transits and Oscillations of stars} (PLATO), to be launched in 2024 \citep{plato2014}.

The surface distribution of stellar magnetic activity, in principle, can be measured with asteroseismology. 
\citet{Gizon2002} investigated the challenge of spatially resolving surface magnetic activity in other stars. He concluded that it would be possible to discriminate between a polar cap distribution and equatorial band distributions of activity, but only for stars with an inclination angle higher than $40^\circ$, activity stronger than the Sun, and rotation strong enough that the individual $m$-components in the acoustic spectra could be resolved.
In a later study, using an empirical model calibrated to the Sun, \citet{Chaplin2007a} tested the ability to measure, with asteroseismology, the extension of a latitudinal activity band distribution in Sun-like stars, finding that the best prospects for detection are for stars with magnetic activity concentrated at low latitudes.
More recently \citet{Santos2012} studied the indirect (nonmagnetic) effects on radial acoustic oscillations induced by the changes in the stellar stratification due to starspots. 
In the case of the Sun they found that the frequency changes are too small (by two orders of magnitude) to explain the observed shifts.  They concluded that the indirect effects of starspots on the stellar stratification cannot be responsible for the observed changes in the acoustic oscillations, at least for a star with a solar-like level of activity.

\subsection{Our approach: Time-domain numerical simulations of waves and starspot}

In this work we extend the investigation to starspots in stars with a higher level of activity: those stars likely have starspots with larger surface coverage, thus implying considerable changes in the internal stratification (\eg,~in the sound speed). 
We also explore the possibility of identifying starspots through asteroseismic observations of highly spotted stars.
We address two main questions: what are the changes in the mode frequencies and amplitudes in such stars? And, what is the starspots seismic signature in synthetic light curves?

We simplify the problem by focusing on the interaction between the acoustic wavefield and a single starspot with a polar geometry \citep[such a configuration is compatible with Doppler observations of a young solar analogue with strong activity,][]{Marsden2005}.
For the starspot we model only the indirect changes induced in the sound speed in the stellar interior.
These changes in principle are not small, therefore, unlike in \citet{Santos2012}, a perturbative approach may not be appropriate, and we need to use direct numerical simulations
accounting for the full 3D structure of the star.
Direct 3D numerical simulations also provide synthetic observations, and therefore are a powerful tool to characterize how the observable quantities change depending on the 3D stellar background.

The remainder of this work is organized as follows:
in the next section we introduce the \textit{GLobal Acoustic Spherical Simulator} (GLASS) code, which extends the code developed by \citet{Hanasoge2006} to include treatment of the center of the star. GLASS simulates linear acoustic waves propagating through the full 3D stellar interior, in the time-domain.
In Sects. \ref{sec:frequencies} and \ref{sec:eigenxi} we describe the spot model employed, analyze the effect of the spot on acoustic modes, and discuss the changes to the eigenmodes in the nonlinear regime.
Section \ref{sec:light curve} highlights the signature of the starspot in the power spectra of the modeled light curves. Finally in Sect. \ref{sec:discussion} we summarize our findings and discuss the observational consequences.

\section{Numerical method }

\subsection{Time-domain pseudo-spectral simulations in spherical geometry}

The GLASS code solves the 3D linearized hydrodynamic equations in a spherical domain containing the full star, from the center up to the stellar surface. 
The use of the linear approximation is justified since acoustic wave perturbations in the Sun and solar\textendash{like} stars have much smaller amplitudes compared to the stellar background quantities (\eg,~velocity perturbations at the surface are $<20~\mathrm{cm/s}$,  four orders of magnitude smaller than the local sound speed, see \citealt{libbrecht1988}).
For the stellar model we considered a spherically symmetric static equilibrium described by Model S \citep{JCD1996} stabilized against convection \citep{Papini2014} and including the photosphere up to $\rmR=1.0007~\rmR_\odot$, $\rmR_\odot$ being the solar radius: that is our quiet Sun (QS) background model.
We then added the starspot model to the background.
We stress here that the spot model must not reintroduce convective instabilities and must fulfill the condition of hydrostatic equilibrium.
We also neglected the perturbation to the gravitational potential \citep[the so-called Cowling approximation,][]{cowling1941}, which reduces the order of the equations by 2.
The use of the Cowling approximation is not, in general, appropriate when considering a full stellar model including the center. However, since here we are only concerned with the changes in the modes of oscillation caused by perturbations to the stellar model in the near-surface layers, we expect the Cowling approximation to be reasonable.
Finally we assumed that the waves propagate adiabatically inside the star.
With these assumptions, the main equations are
\begin{equation}
  \label{eq:dtxi}
  \PD{\bxi}{t} = \bv,
\end{equation}
\begin{equation}
  \label{eq:dtv}
  \PD{\bv}{t} = -\frac{\GRAD{p}}{\rho_0} -\frac{\rho}{\rho_0}g_0\be_r - \Gamma(r)\bv,
\end{equation}
where $\bxi(\br,t)$ is the vector displacement of the wavefield, $\bv(\br,t)$ is the Eulerian velocity perturbation, $g_0(r)\be_r$ and $\rho_0(r)$ are the unsigned (radially directed) gravitational acceleration and density of the stellar background, and $\br$ and $\be_r$ are the position vector and the unit vector in the radial direction. Here
$\Gamma(r)$ is a sponge\textendash{like} damping term that prevents spurious waves reflection at the upper boundary \citep{Hanasoge2006}. 
The symbols $\rho(\br,t)$ and $p(\br,t)$ are the Eulerian perturbations to density and pressure, both functions of $\bxi$ by means of the linearized continuity and adiabatic equations
\begin{equation}
  \label{eq:rho}
  \rho= -\DIV (\rho_0\bxi), 
\end{equation}
\begin{equation}
  \label{eq:p}
  p = -\bxi \cdot \GRAD p_0 - \rho_0 \left ( c_0^2 + \Delta c^2 \right ) \DIV \bxi,
\end{equation}
where $\Delta c^2$ is the change in the squared sound speed induced by the starspot, $p_0$ and $c_0^2=\gamma_0 p_0/\rho_0$ are respectively the pressure and the square of the adiabatic sound speed of the stellar background, and $\gamma_0$ is the first adiabatic exponent. 
The wavefield is excited by setting an initial condition for the displacement $\bxi$, then the code performs the temporal integration for $\bxi$ and $\bv$, while the other relevant quantities are computed at each timestep.

GLASS employs a pseudo-spectral scheme, which uses spherical harmonic (SH) decomposition on spherical surfaces to compute the horizontal derivatives, and a sixth-order tridiagonal compact scheme \citep{Lele1992} for radial derivatives.
In SH space the spectral components are identified by
the angular degree $\ell$ and the azimuthal order $m$.
Temporal integration is performed by means of a five-stage second-order low dissipation and dispersion Runge-Kutta (LDDRK) scheme \citep{Hu1996}.
For a more detailed description of the code see \citet{Hanasoge2006}.
The grid size is chosen according to the desired resolution: the maximum value $\ell_{\max}$ of $\ell$ in the SH transform sets the minimum number of latitudinal and longitudinal grid points ($n_\mathrm{lon}=2n_\mathrm{lat}\geq3\ell_{\max}$), while in the radial direction we adopted a grid with a step size constant in acoustic depth, small enough to resolve the background model and the shortest wavelength among the modes of interest.
Upper boundary conditions are set by imposing
a vanishing Lagrangian perturbation of the pressure (which translates into the condition $\DIV\bxi=0$ at $r=\rmR$), while at the center of the star we prescribe regularity conditions, as we will now explain.

\subsection{Extending the simulation to the center of the star} 

Particular care must be taken when considering the center of the star. 
This point is a geometrical singularity in spherical coordinates,
therefore, when calculating the radial derivatives for $\xi_r$ and $p$, boundary conditions at the center are imposed by considering the asymptotic behavior of radial displacement and pressure in SH space.
The spectral components $\xi_{r,\ell m}$ and $p_{\ell m}$ behave like $r^{\ell-1}$ and $r^\ell$  respectively when $r\rightarrow 0$ (see, \eg, \citet{Unno1979}).
We enforce these conditions at the center by requiring that the radial derivatives of  
$\xi_{r,\ell m}/r^{\ell-2}$ and $p_{\ell m}/r^{\ell-1}$
vanish at the center. We then obtain the original derivatives $\PD{\xi_{r,\ell m}}{r}$ and $\PD{p_{\ell m}}{r}$ by means of algebraic formulas.
This procedure also ensures that the numerical accuracy of the compact scheme is preserved near the center.

The horizontal spatial resolution increases with depth, owing to the clustering of the grid points, while the radial resolution remains roughly constant.
As a consequence, a numerical instability appears in those spectral components for which the horizontal wavenumber $k_{h,\ell}=\sqrt{\ell(\ell+1)}/r$ exceeds the 
Nyquist wavenumber $k_{r,\max}=\pi/\Delta r$ in the radial direction, $\Delta r$ being the radial grid step.
We remove this instability by employing a spectral filter that, at each radial grid point $r_j$, sets to zero all the spectral components satisfying $ k_{h,\ell}(r_j) > k_{r,\text{max}}$.
For a grid with constant $\Delta r$ this translates to the condition
\begin{gather}
  j < j_\ell = \frac{\sqrt{\ell(\ell+1)}}{\pi} \quad(\;+\; \text{offset}),
\end{gather}
where $j$ identifies the $j$-th grid point from the center.
A safety offset parameter is also implemented (we found that 2 is the smallest value that  removes the instability).
Finally, a fourth-order tridiagonal compact low-pass filter \citep{Lele1992} is applied in the radial direction, to avoid spectral blocking \citep{Hanasoge2007} and smooth the discontinuity in $r$ caused by the spectral filter.
The physical solutions are not affected by the filtering; radial modes (the only modes that propagate through the center of the star) are not filtered, and all the other modes are already evanescent at the points $r_j <r_{j_\ell}$ (the lower turning point for an $\ell=1$ mode at $5~\mHz$ is $0.03~\rmR_\odot$, \ie, the tenth gridpoint in the grid we employ, while $j_{\ell=1} = 2$). However the combined action of the filter with the time evolution scheme introduces damping in the wavefield, with an
exponential dependence on frequency. This results in a lifetime of $\sim 50~\days$ for waves with frequency at $2~\mHz$ and down to $\sim 3~\days$ at $5~\mHz$ for the simulation setup we used, which employs a radial grid with $n_\mathrm{rad}=800$ grid points and a time step of $1 ~\mathrm{s}$.

\subsection{Validation: Comparison with ADIPLS normal modes}

We validated GLASS through independent calculations of the theoretical modes of oscillations for our QS model.
These modes are uniquely identified by three quantum numbers: the radial order $n$, the angular degree $\ell$, and the azimuthal order $m$.
With the ADIPLS software package \citep{Christensen-Dalsgaard2008}, we computed the eigenfrequencies $\omega_{0,n\ell}$ (which are degenerate in $m$ in the QS model) and eigenfunctions $\bxi_{0,n\ell m}$ of all the acoustic modes in the range $3<n<37$ and up to $\ell=2$.
We then used the eigenfunctions given by ADIPLS to excite the initial wavefield displacement in GLASS.
Test simulations were run in a temporal window of 10 days (solar time), using different initial conditions (by exciting either one or a few eigenmodes at the same time) and grid sizes. We also performed one simulation with all the modes excited and in a temporal window of 70 days, to use as reference for our study.

In all the tests performed, the extracted eigenfrequencies and eigenfunctions were compared with ADIPLS solutions and showed good agreement: the difference between ADIPLS and GLASS for the eigenfrequencies was below $0.16~\mu\Hz$ (\ie, the frequency resolution of the simulations) and the maximum difference for the eigenfunctions was $\sim0.1\%$, after the damping was accounted for.

\section{Frequency shifts: Nonlinear dependence on perturbation amplitude}
\label{sec:frequencies}

\subsection{Spot model: Perturbation in sound speed}
We modeled the changes to the stratification caused by a starspot as a positive change  $\Delta c^2$ in the squared sound speed, while the density and pressure were unchanged. This guaranteed the compatibility with the stabilization method used for the background model (for which a decrease in the sound speed would reintroduce convective instabilities).
We defined the change in the sound speed as
\begin{equation}
\label{eq:deltagammaspot}
 {\Delta c^2(r,\theta)} = \mathrm{\epsilon} c_0^2(r)   f(r;r_c) g(\theta), 
 \end{equation}
with positive amplitude $\epsilon$, a radial profile
\begin{equation*}
  f(r;r_c)=
 \frac{1}{2}\left [\cos\left ({|r-r_c|}/{\sigma}\right )+1\right ]\exp{\left [-{(r-r_c)^2}/({2\sigma^2})\right ]}
\end{equation*}
for $|r-r_c| < \pi{\sigma}$ and zero otherwise,
and with a latitudinal profile
\begin{equation*}
g(\theta)=
\frac{1}{2}\left [\cos\left (\kappa \theta\right )+1\right ]  
\end{equation*}
for $\kappa \theta < \pi$ and zero otherwise,
where $\theta$ is the colatitude.
The functions $f$ and $g$ have continuous derivatives everywhere and define a spot located at the north pole, with a Gaussian profile in radius multiplied by a raised cosine and a raised cosine profile in latitude.
For the study the vertical and horizontal size of the spot were fixed by setting $\sigma=0.01~\rmR_\odot$ and $\kappa=2.4$, and we varied the depth $\rmR_\odot-r_c$ and the amplitude $\epsilon$. 
We note that the choice of the coordinate system  here is completely arbitrary, since we are studying the perturbation to a spherically symmetric (\eg, nonrotating) background, 
therefore the results of this work can be translated to a spot located at any point at the surface, via a rotation of the coordinate system.

\subsection{Linear theory}
We first discuss the effect of small perturbations, in order to determine for which parameter range the frequency shifts falls in the linear regime.
We consider the normal mode solutions $\bxi_0(\br,t)=\bxi_{0,n\ell m}(\br)e^{-i\omega_{0,n\ell} t}$ of our reference QS model, 
 solving the wave equation 
\begin{equation}
 \rho_0\omega_{0,n\ell}^2\bxi_0=-\nabla(c_0^2\rho_0 \DIV{\bxi_0}) - \nabla(\bxi_0\cdot\nabla{p_0}) + g_0 \be_r\DIV{(\rho_0\bxi_0)} = \mH\bxi_0
\end{equation}
that can be derived from Eqs. (\ref{eq:dtxi}-\ref{eq:p}).
 Since the reference background is spherically symmetric, the solutions are degenerate in $m$.
We consider now the perturbation $\Delta\mH$ to the wave operator $\mH$ due to a generic change in the sound speed (while retaining pressure and density constant), given by
\begin{equation}
 \Delta\mH\bxi_0= - \nabla \left ( \Delta c^2\rho_0 \DIV{\bxi_0}\right );
\end{equation}
the expression for the linear frequency shift (see, \eg, \citealt{2010Asteroseismo}) is
\begin{equation}
\label{eq:linearshift}
 \frac{ \Delta\omega_{n\ell m}}{\omega_{0,n\ell}}= \frac{\int_\odot\Delta c^2  (\Div{\vet{\xi}_0^*})(\Div{\vet{\xi}_0})\rho_0\df{V}}{2\omega_{0,n\ell}^2\int_\odot \|\bxi_0\|^2 \rho_0 \mathrm{d}V}, 
\end{equation}
where integration is performed over the entire volume of the star. 
Both the surface integral and the perturbation to the mode inertia do not contribute to the change in $\Delta\omega_{n\ell m}$ in this case:
the surface integral is zero because of the surface boundary conditions imposed for the pressure, and there is no contribution from the perturbation to the mode inertia, since the density of the background is unchanged.
Finally, by exploiting the separable form of the polar spot profile (\ref{eq:deltagammaspot}), we may write the frequency shift in terms of a product of separate integrals for the horizontal and radial coordinates.
It is worth noting that, unlike the case of splitting induced by rotation, the frequency shift depends only on $|m|$ and therefore modes with the same $|m|$ are still degenerate. This holds true also in the nonlinear case.

\subsection{Numerical simulations}
\label{sec:numericalsim}
\subsubsection{Initial conditions: $\delta$-function source}
We performed our study in the parameter range $ 0.01 \le \epsilon \le 1$ and $ 0.97~\rmR_\odot \le r_c \le 1~\rmR_\odot$.
The simulations run for 70 days (solar time) to reach the desired accuracy of $\sim 0.16~\mu{\Hz}$ in the frequency domain. 
Wavefield displacement and velocity records were taken with a cadence of 60 seconds (solar time), mimicking the usual cadence of helio- and asteroseismic observations and in order to have a Nyquist frequency of $\sim 8.3~\mHz$ (above the maximum acoustic cutoff frequency of $\sim5.3~\mHz$). 
Starting from the ADIPLS eigenfunctions, we set the initial conditions for the wavefield displacement as 
\begin{equation}
\label{eq:IC}
 \bxi(\br,t=0)=\sum_{n=3}^{37} \sum_{\ell=0}^{2} \sum_{m=-\ell}^{\ell}
  \bxi_{0,n\ell m}(\br),
\end{equation}
which excites all the modes from $3<n<37$ and up to $\ell=2$. The initial velocity $\bv$ is set to zero. We note here that all the modes were excited with the same phase.
\subsubsection{Nonlinear frequency shifts}

The simulated frequencies have been extracted by taking the SH transform coefficient ${p}_{\ell m}(r_0,t)$ of the wavefield pressure $p(r_0,\theta,\phi,t)$ at each timestep and at $r_0=\rmR_\odot+200~\mbox{km}$ above the surface. This was followed by a Fourier transform in time to obtain the field $\tilde{p}_{\ell m}(r_0,\omega)$ and then the power spectrum $P_{\ell m}(\omega)= | \tilde{p}_{\ell m}(r_0,\omega)|^2$. 
Finally we divided each $(\ell,m)$ spectrum in chunks with size of $80~\mu\Hz$ and centered on the peak closest to the original ADIPLS mode frequency. A least-squares Lorentzian fit was applied to extract frequency, amplitude, and half width at half maximum (HWHM) of each mode.
\begin{figure}
 \includegraphics[width=1.\columnwidth]{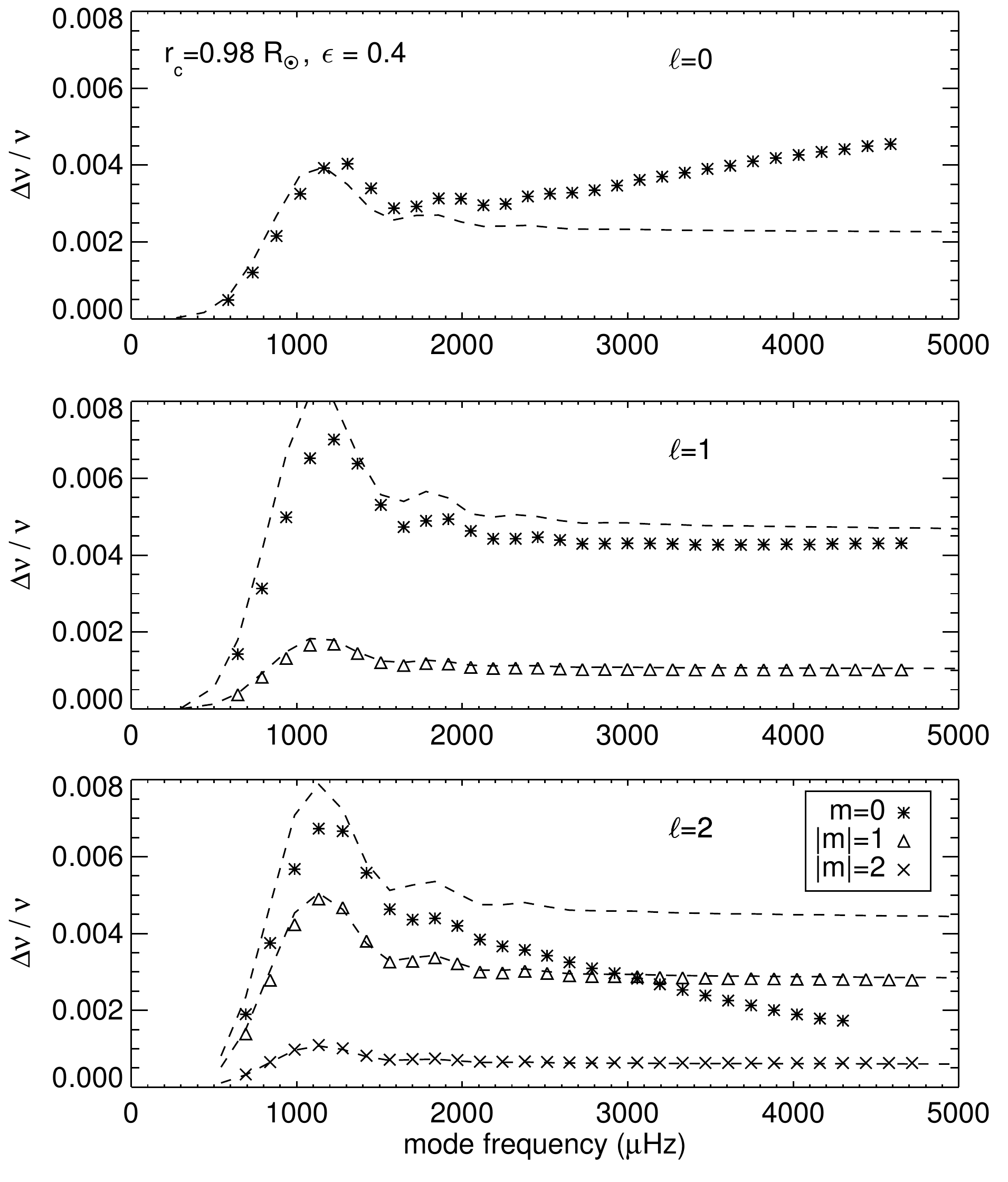} 
 \caption[frequency shifts for all modes]{\begin{footnotesize}
     Relative frequency shifts $\Delta\nu/\nu$ vs. mode frequency $\nu$ for modes with $\ell=0$ (upper panel), $\ell=1$ (middle panel) and $\ell=2$ (bottom panel), for a polar spot model relative to QS model with $r_c=0.98 ~\rmR_\odot$ and $\epsilon=0.4$. Dashed lines represent linear frequency shifts given by Eq. (\ref{eq:linearshift}). Fitted frequencies from numerical simulations (asterisks for $m=0$, triangles for $|m|=1$, and crosses for $|m|=2$) show the nonlinear behavior of the shifts for $m=0$ modes (asterisks).
     We note that $m=0$ and $|m|=1$ mode frequencies for $\ell=2$ cross at $\nu\simeq 3060 ~\mu\Hz$.
 \end{footnotesize}}
 \label{fig:manymodes098}
\end{figure}

\begin{figure}
 \includegraphics[width=1.\columnwidth]{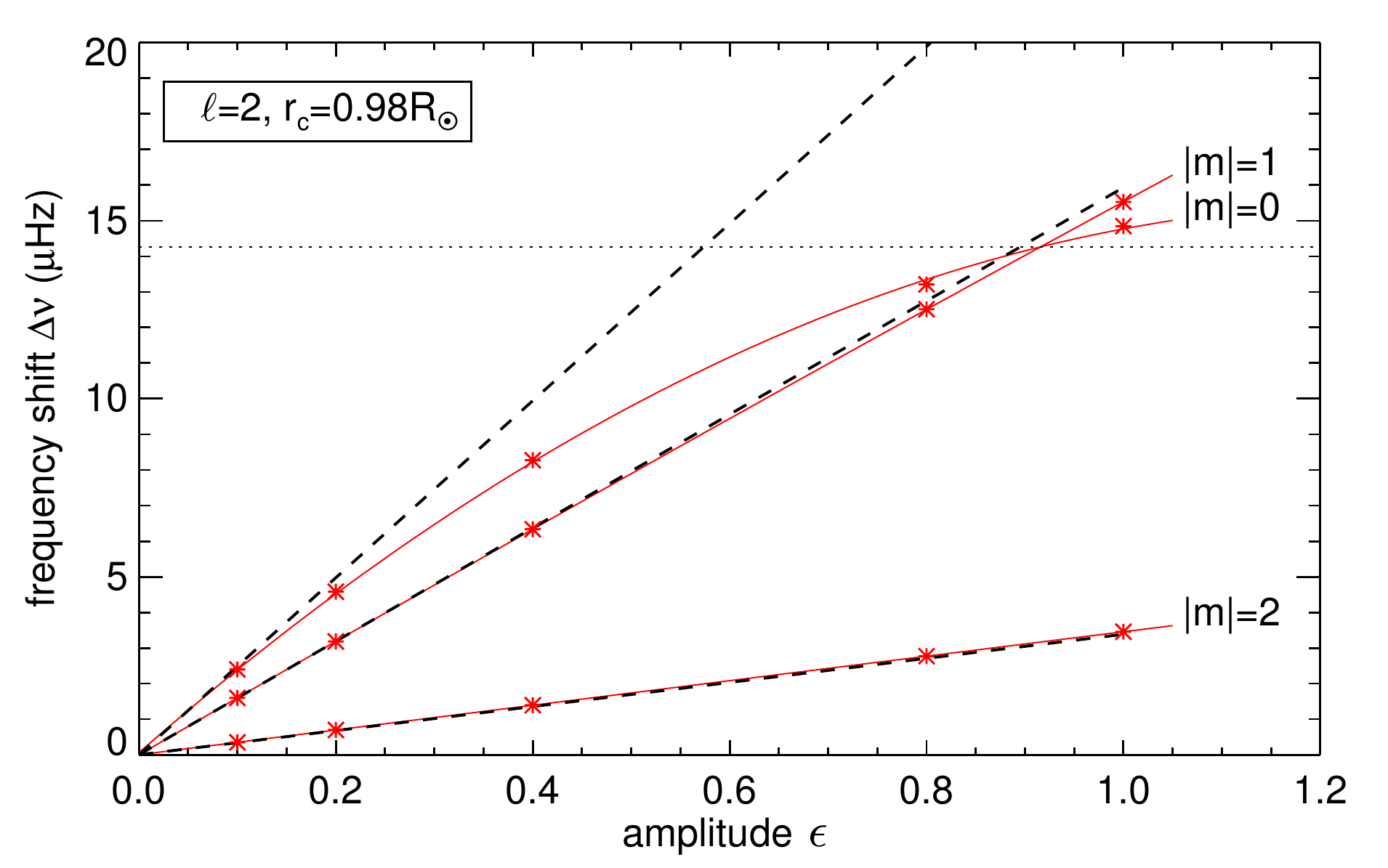}
 \caption[frequency shifts vs. $\epsilon$ for $\ell=2$, $n=12$]{\begin{footnotesize}
     Mode frequency shifts $\Delta\nu$ vs. spot amplitude $\epsilon$ for modes with $\ell=2$, $n=12$ and $|m|=0,1,2$
     in the case of a polar spot model with $r_c=0.98$. The QS eigenfrequency for these modes is $\nu_0=1970.50 \pm 0.16~\mu\Hz$. Dashed lines represent linear frequency shifts given by Eq. (\ref{eq:linearshift}), red asterisks are frequencies shifts from numerical simulations. Solid red lines represent  parabolic fits to the shifts.
     The horizontal dotted line shows the crossing of the frequencies between the $m=0$ and $|m|=1$ modes at $\epsilon\simeq0.92$.
 \end{footnotesize}}
 \label{fig:dnuvseps098}
\end{figure}

Figures \ref{fig:manymodes098} and \ref{fig:dnuvseps098} show selected results obtained for the frequency shifts $\Delta\nu=\Delta\omega/2\pi$ induced by the spot with respect to QS in the case of a polar spot located at $r_c=0.98~\rmR_\odot$ (in the following we will always show plots related to simulations performed for this value of $r_c$).
Figure \ref{fig:manymodes098} shows the relative frequency shifts $\Delta\nu/\nu $ as a function of mode frequency $\nu=\omega/2\pi$ for all the modes we excited, extracted from a simulation with a spot of amplitude $\epsilon=0.4$. 
The linear frequency shifts calculated from Eq. (\ref{eq:linearshift}) reproduce the behavior of nonaxisymmetric modes ($m\neq0$). 
However for axisymmetric modes the frequency shifts deviate from the theoretical linear shifts, but with a different behavior depending on $\ell$, the most interesting being the case $\ell=2$ (bottom panel). 
Here the relative frequency shifts for $m=0$ modes are smaller than predicted, and even decrease with increasing mode frequency, while $\Delta\nu/\nu$ remains roughly constant for the $|m|=1 \text{ and }2$ modes. As a consequence, at a frequency of $\sim 3000~\mu\Hz$ the $\Delta\nu/\nu$ for $m=0$ modes cross the relative frequency shifts of the $m=1$ modes. Above this crossing frequency the $m=1$ modes have the largest frequency shifts.
Results from the other simulations performed shown that this crossing frequency decreases either when the amplitude $\epsilon$ increases (as seen, \eg, in Fig. \ref{fig:dnuvseps098}) or when $r_c$ is moved toward the surface, the latter indicating a stronger nonlinear response of the system to smaller surface changes than to bigger changes buried deeper in the convection zone.

Figure \ref{fig:dnuvseps098} shows the simulated frequency shifts
$\Delta\nu$ against spot amplitudes $\epsilon$ and
for all the modes with $\ell=2,n=12$. These modes in the reference QS model are degenerate with respect to the azimuthal order $m$, with a frequency $\nu_0=1970.50 \pm 0.16~\mu\Hz$. 
Again linear theory successfully reproduces the frequency shift for nonaxisymmetric modes, shifts for $|m|=1$ however start to deviate from linear behavior at $\epsilon \simeq 0.8$.
The frequency shifts for $m=0$ modes on the contrary are nonlinear already at $\epsilon$ values of $\sim 0.2$, with $\Delta\nu$ values smaller than predicted by linear theory.
A parabolic fit is able to model the shift, thus indicating that in this case a second order perturbative correction could recover the actual frequencies.
The figure shows also the crossing frequency (horizontal dotted line) matching the mode frequency at $\epsilon \simeq 0.92$, with a value of about $1985~\mu\Hz$ (corresponding to a frequency shift $\Delta\nu\simeq14.5~\mu\Hz$), decreased by $\sim 1000~\mu\Hz $ with respect to the case with $\epsilon=0.4$, as already noted above.

\section{Perturbations to the eigenfunctions}
\label{sec:eigenxi}
The polar spot introduces a nonspherically-symmetric perturbation to the background model. As a consequence the eigenfunctions depart from their original horizontal dependence and get mixed with other spherical harmonics. The extent of the mixing depends on the amplitude of the perturbation. Because of the axisymmetric profile of the spot there is no mixing between spherical harmonics with different $m$.
We investigate the effect of the mixing in the eigenfunctions by writing the radial displacement eigenfunction of a given mode for the model with the spot as
\begin{equation}
 \label{eq:xir_thtar}
 \xi_{r}^{n\tilde\ell m}(\br,t)=
 e^{-i\omega_{n \tilde{\ell}m} t}
 \sum_{\ell=0}^{\ell_{\max}} a_{\ell}^{n\tilde{\ell}m}(r)\Ylm(\theta,\phi)
 =
 \xi_r^{n \tilde{\ell}m}(r,\theta)e^{im\phi} e^{-i\omega_{n \tilde{\ell}m} t}
\end{equation}
with the spherical harmonics $\Ylm$ given by
\begin{equation*}
 \Ylm(\theta,\phi)=\sqrt{\frac{(2\ell+1)}{4\pi}\frac{(\ell-m)!}{(\ell+m)!} } P_\ell^m(\cos\theta) e^{im\phi},
\end{equation*}
where $\ell_{\max}$ gives the spectral resolution of the SH transform
and where we labeled with $\tilde{\ell}$ the new mixed eigenmode originally represented by a pure $Y_{\tilde{\ell}}^m $ in the QS model. 
Here $\xi_r^{n \tilde{\ell}m}(r,\theta)$ is the meridional profile of $\xi_{r}^{n\tilde\ell m}$, expressed through a truncated series of Legendre polynomials (\ie, the latitudinal components of the $\Ylm$)
\begin{equation}
 \xi_r^{n \tilde{\ell}m}(r,\theta) =\sum_{\ell=0}^{\ell_{\max}} a_{\ell}^{n \tilde{\ell}m}(r) \sqrt{\frac{(2\ell+1)}{4\pi}\frac{(\ell-m)!}{(\ell+m)!} } P_\ell^m(\cos\theta).
\end{equation}
The amplitude coefficients $a_{\ell}^{n \tilde{\ell}m}(r)$ with $\ell \neq\tilde{\ell}$ give the degree of mixing with $\Ylm$ of the $n\tilde{\ell}m$ mode at the radial position $r$.
\begin{figure}
 \includegraphics[width=1\columnwidth]{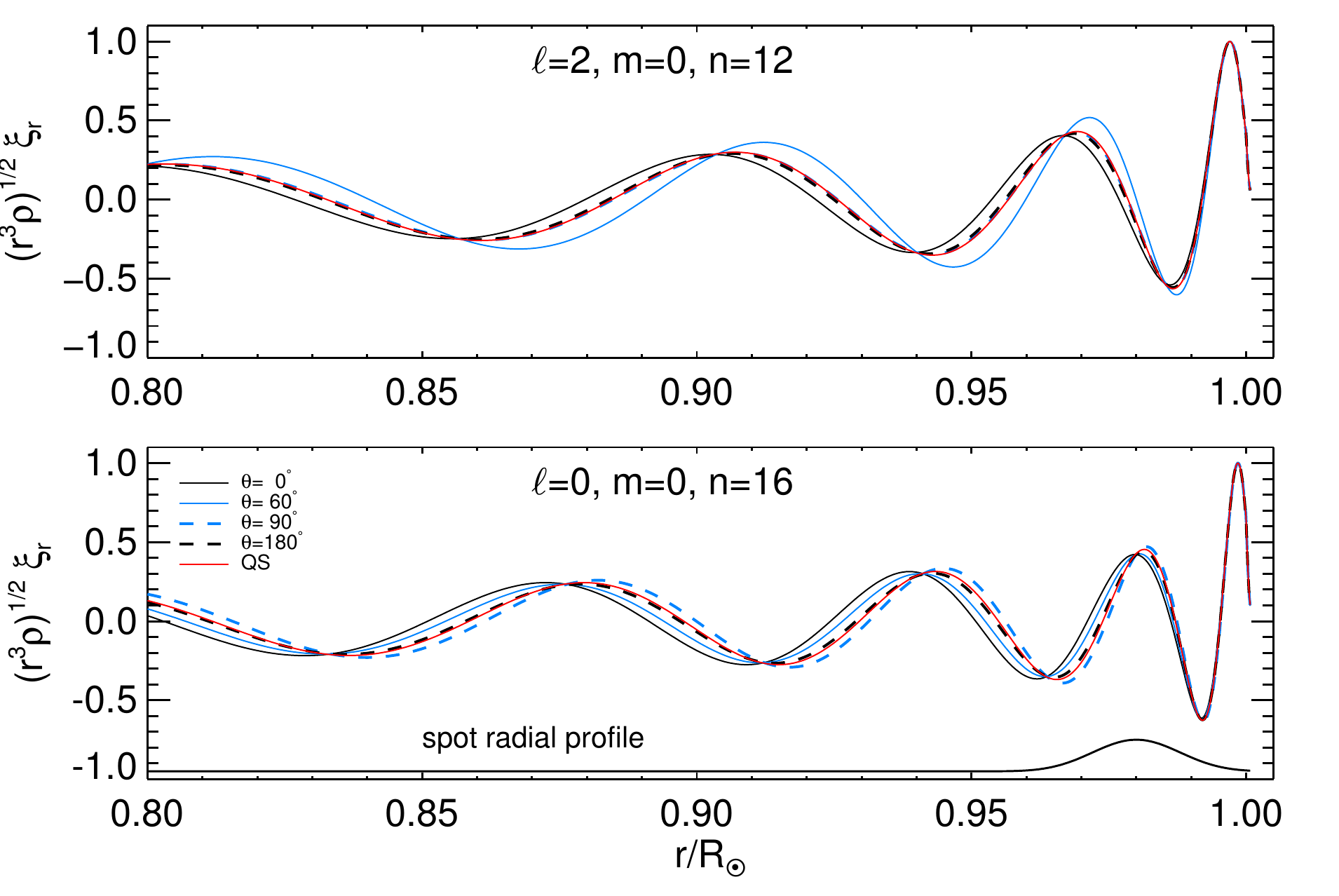}
 
 \caption[Radial displacement eigenfunction]{\begin{footnotesize}
     Normalized cuts at different colatitudes $\theta$ of the radial displacement eigenfunction $(r^3\rho)^{1/2}\xi_r(r,\theta)$ at the center of the spot ($\theta=0$, solid black line), at the edge ($\theta=60^\circ$, solid blue line), and outside the spot (dashed lines),
     of the $\tilde\ell=2$, $m=0$, $n=12$ mode (top) and $\tilde\ell=0$, $m=0$, $n=16$ mode (bottom), for a polar spot with $r_c=0.98~\rmR_\odot$ and $\epsilon=0.4$. 
     The solid red line is the corresponding $\xi_r$ from ADIPLS.
     The solid black line in the bottom panel beneath the eigenfunctions shows the radial profile $f$ of the spot.
     The blue and black dashed lines in the top panel completely overlap.
 \end{footnotesize}}
 \label{fig:eigenxir098}
\end{figure}

Figure \ref{fig:eigenxir098} shows radial cuts taken at different colatitudes $\theta$ (\ie, at different angular distances from the center of the spot) of the scaled radial displacement eigenfunction $(r^3\rho)^{1/2}\xi_r^{n\tilde{\ell}m}(r,\theta)$, for a $\tilde\ell=2$, $m=0$, $n=12$ (top panel) and for a radial mode with $n=16$ (bottom panel), for a spot with $\epsilon=0.4$ and $r_c=0.98~\rmR_\odot$. Each cut has been renormalized to its maximum. 
In the case of the quadrupolar mode the phase shift of $\xi_r$ inside the spot with respect to the QS solution increases from the center, and reaches its maximum at the edge of the spot (\ie, for $\theta=60^\circ$). The shift then decreases, and $\hat\xi_r$ smoothly matches the QS eigenfunction at $\theta=90^\circ$.
The phase shift has a different behavior in the case of the radial mode with $n=16$,
it is maximum at the center of the spot and then approaches zero at the antipodes ($\theta=180^\circ$). 
This is in agreement with what was found by \citet{Santos2012}, except that in our case the radial profile of the eigenfunction does not match the QS eigenfunction at the spot edge.

\section{Synthetic power spectra for a polar spot}
\label{sec:light curve}

\subsection{Synthetic light curves}
Mode mixing affects observed light curves, as was already pointed out by \citet{Dziembowski1996} and \citet{Cunha2000} in the case of RoAp stars, since it changes the expected mode visibilities.
The production of a realistic synthetic light curve is a nontrivial task, which requires modeling all the contributions to the emergent intensity at the photosphere in the observed wavelength range. Here we only model the contribution of the photospheric pressure perturbations to the intensity fluctuations $I$ induced by the acoustic wavefield. This in principle requires an explicit relation between the mode displacement and $I(\theta',\phi',t)$ at the stellar surface, which is rather complicated in general (see \citealt{Toutain1993}).

For the sake of simplicity we assume that $I(\theta,\phi,t)$ is proportional to the Eulerian pressure perturbation $p(r_0,\theta,\phi,t)$  measured at $r_0=R_\odot + 200~\text{km}$. We then express $I$ in the frequency domain as
\begin{equation}
\label{eq:intensity1}
\tilde I(\theta,\phi,\omega) \propto \tilde p(r_0,\theta,\phi,\omega)= \sum_{\ell=0}^{\ell_{\max}}\sum_{m=-\ell}^{\ell}  \tilde{p}_{\ell m}(r_0,\omega)  
 \Ylm(\theta,\phi),
\end{equation}
where ${p}_{\ell m}$ are the SH coefficients of the pressure, obtained as described in Section \ref{sec:numericalsim} and containing all the contributions to the wavefield in the spectral component $(\ell,m)$, including the mixing from other $\ell\neq\tilde\ell$ modes.
We can explicitly quantify the mixing of a single mode by writing (see also Eq. (\ref{eq:xir_thtar}))
\begin{equation}
\label{eq:contamination}
\tilde p(r_0,\theta,\phi,\omega_{n\tilde\ell m})=\sum_{\ell=0}^{\ell_{\max}}  
 \tilde{p}_{\ell m} (r_0,\omega_{n\tilde\ell m})\Ylm(\theta,\phi),
\end{equation}
where $\omega_{n\tilde\ell m}$ is the mode frequency in the presence of the spot.
The values of $\tilde{p}_{\ell m}(r_0,\omega_{n\tilde\ell m})$ with $\ell\neq \tilde\ell$ give the degree of the mixing.

\begin{figure}
 \includegraphics[width=1.\columnwidth]{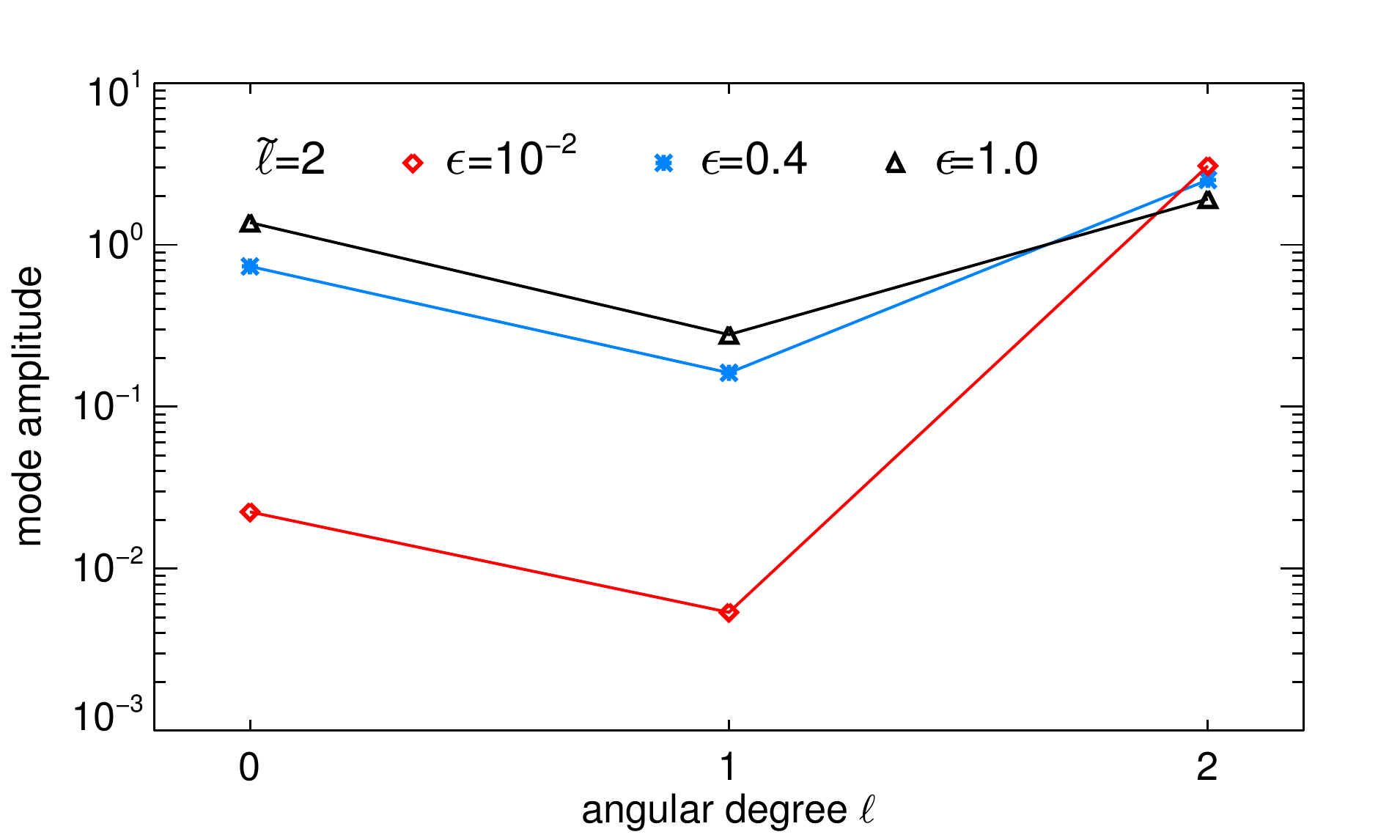}
 \caption[Contamination]{\begin{footnotesize}
     Amplitude coefficients $|\tilde{p}_{\ell m}(r_0,\omega_{n\tilde\ell m})|$ of equation (\ref{eq:contamination}) showing the mixing of the $\tilde\ell=2,n=12,m=0$ mode for three simulations with spot amplitude $\epsilon=0.01$ (red squares), $0.4$ (blue asterisks), and $1.0$ (black triangles). 
 \end{footnotesize}}
 \label{fig:amplitudes}
\end{figure}

Figure \ref{fig:amplitudes} shows the coefficients $|\tilde{p}_{\ell m}(r_0,\omega_{n\tilde\ell m})|$ of expansion (\ref{eq:contamination}) for the $n=12,\,\tilde\ell=2,\,m=0$ mode for three different values of $\epsilon$. 
Mixing in the $\epsilon=0.01$ case is almost absent. 
For $\epsilon=0.4$ the mixing contribution from the dipole component is  negligible (one order of magnitude smaller), while that from the radial component is $30\%$ of the leading mode. The mixing increases significantly for $\epsilon=1$,
reaching $15\%$ for $\ell=1$ and $72\%$ for $\ell=0$, thus resulting in extreme distortion of the eigenfunctions for that mode.

\subsection{Example power spectrum for $\ell=2$}
\begin{figure}
  \includegraphics[width=1.\columnwidth]{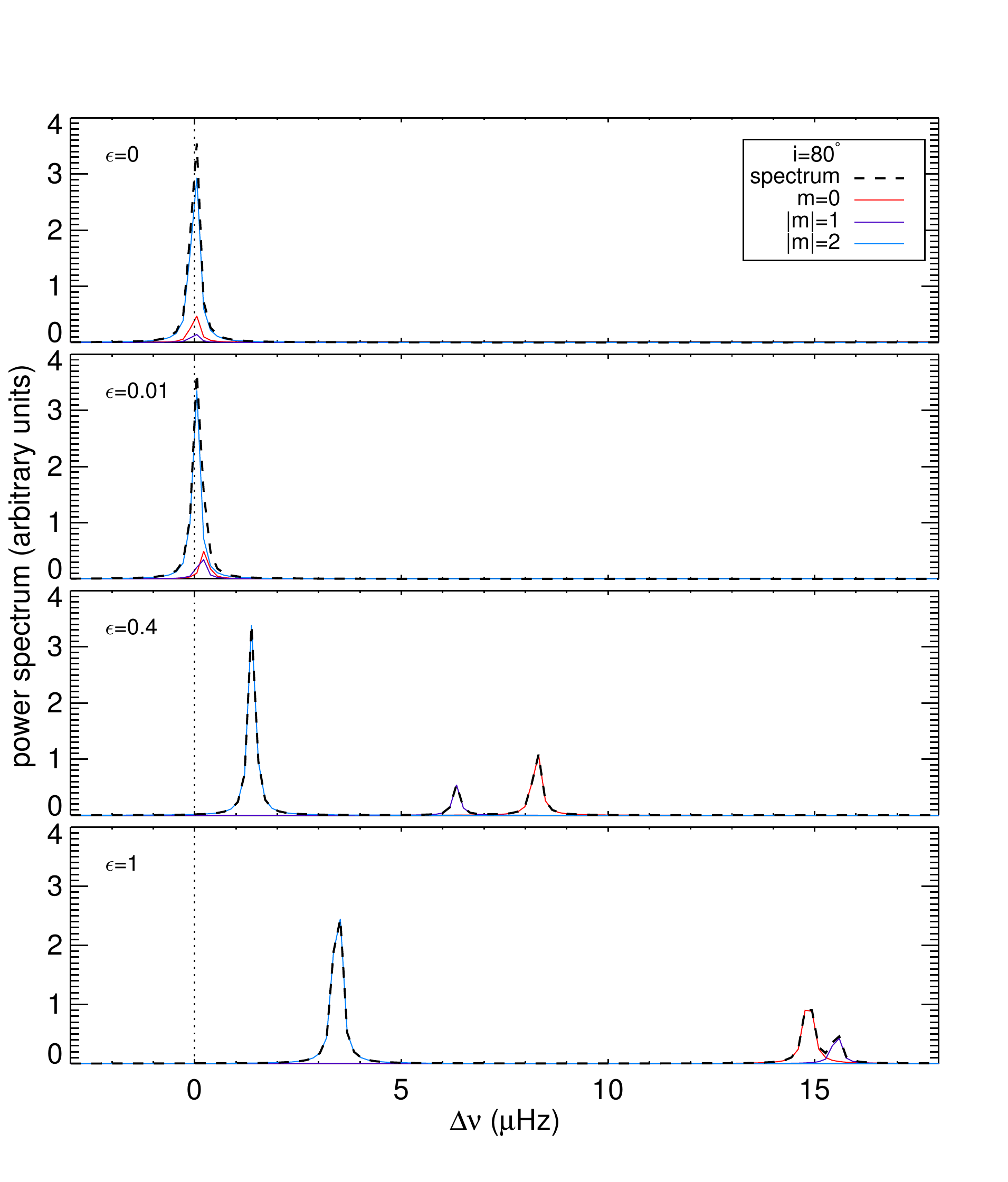}
 \caption[Power spectrum for $i=80^\circ$]{\begin{footnotesize}
      Acoustic power vs. $\Delta\nu$ for $\tilde\ell=2$, $n=12$ and $m=0$ (red) $,\pm 1$(violet), and $\pm 2$ (blue) at an inclination angle $i=80^\circ$, for QS ($\epsilon=0$, upper panel) and for a spot located at a depth of $0.98~\rmR_\odot$ and with $\epsilon=0.01$ (second upper panel), $0.4$ (third upper panel), and $1.0$ (bottom panel). The black dashed lines represent the spectra resulting from the sum of all the $m$ components.
      Because of the initial conditions in Eq. (\ref{eq:IC}), the observed acoustic power in the $|m|>0$ modes is twice the power that would be observed in the case of stochastic excitation.
 \end{footnotesize}}
 \label{fig:lightspectrum}
\end{figure}

As an example we consider now four synthetic power spectra $P(\omega)$ of a star with an inclination angle $i$ of the polar axis with respect to the line of sight of $80^\circ$, for QS and three different  amplitudes with $\epsilon$ values of $0.01$, $0.4$, and $1$, respectively. 
To create the spectra we followed the procedure outlined by \citet{Gizon2003}: starting from the intensity fluctuations approximated by Eq. (\ref{eq:intensity1}), we calculated the disk integrated intensity $\tilde I(\omega)$ in the frequency domain, accounting for projection and limb-darkening effects \citep{Pierce2000}. The resulting power spectra $P(\omega)=|\tilde I(\omega)|^2$ vs.  $\Delta\nu$ are shown in Fig. \ref{fig:lightspectrum} for the $\tilde\ell=2,n=12$ multiplet. 
We note that because of the initial conditions we set (\ie, the choice made in Eq. (\ref{eq:IC}) of using the same phase in exciting all of the modes), the $|m|>0$ peaks in these spectra are twice as high as the peaks in a spectrum of acoustic oscillations resulting from stochastic excitation (that is the case for Sun-like stars).
The $\epsilon=0.01$ case falls in the linear regime: the visibility of the modes is the same as that of pure spherical harmonics (observed amplitude ratios correspond to Fig. 2 of \citet{Gizon2003}, once the residual degeneracy in the $m\neq0$ modes and the choice of the phase in the initial conditions (\ref{eq:IC}) are taken into account) with very little mixing, as already shown in Fig. \ref{fig:amplitudes}. 

The same holds true in the case of $\epsilon=0.4$ and $1$ for nonaxisymmetric modes
(the differences in shape and height between the peaks in the plots occur because the bin size is comparable to the HWHM of the peaks for these modes).
The $m=0$ mode, on the contrary, departs from the linear regime both in frequency shift (which for $\epsilon=1$ becomes even smaller than that one for $|m|=1$) and observed amplitude, as a consequence of the mixing with other spherical harmonics. 

\section{Conclusion}
\label{sec:discussion}

Using 3D linear numerical simulations, we investigated the changes in global acoustic modes with $\ell \leq 2$ induced by a localized sound-speed perturbation with relative amplitude $\epsilon$, mimicking the changes caused by a starspot with a polar cap configuration.

The interaction of the wavefield with a polar spot strongly affects the  axisymmetric modes, which show a nonlinear behavior increasing with $\epsilon$, for $\epsilon \gtrsim 0.2$. 
The frequency shifts for radial modes exceed the shifts predicted by linear theory, while the shifts for the $\ell=2, m=0$ modes are smaller than predicted by linear theory and cross the $|m|=1$ modes at a frequency that decreases with increasing $\epsilon$. 
For modes with $m\neq 0$ linear theory successfully predicts the correct frequencies. 

The nonlinear changes, with $\epsilon$, in the mode frequencies and mode mixing (resulting from the distortion of eigenfunctions)
will play a role in the correct identification of the modes.
Strong mode mixing may also cause $\ell>2$ modes to become visible in the observed spectrum \citep[see, \eg, ][]{Dziembowski1996}.

\begin{acknowledgements}
The authors acknowledge research funding by the Deutsche  Forschungsgemeinschaft (DFG) under the grant SFB~963/1 project A18. 
L.G. acknowledges support from EU FP7 Collaborative Project \textit{Exploitation of Space Data for Innovative Helio- and Asteroseismology} (SPACEINN).
S.H. acknowledges a collaboration with the Max Planck Institute for Solar
System Research through a Max Planck Partner Group established at the Tata Institute of Fundamental Research, Mumbai.
We thank Robert Cameron for useful discussions and Jesper Schou and  Hannah Schunker for useful comments.
\end{acknowledgements}

\bibliography{biblio}

\end{document}